# Application of E-commerce Technologies in Accelerating the Success of SME Operation

Ziad Almtiri[1,2], Shah J. Miah[2], and Nasimul Noman[3]

[1] Department of Management Information Systems, Taif University, Taif, KSA
[2] Newcastle Business School, University of Newcastle, NSW, Australia
[3] School of Information and Physical Sciences, University of Newcastle, NSW, Australia
`ziad_almtiri@outlook.com`, `shah.miah@newcastle.edu.au`,
`nasimul.noman@newcastle .edu.au`

**Abstract.** Application of electronic commerce (e-commerce) technologies have increased notably over the past two decades in different business sectors. In particular, the technologies of B2C operations have significantly improved the productivity of online small businesses such as SMEs. Systematic literature reviews in this domain categorized different benefits but a limited number of studies on SME success from the view of information systems (IS) research exists, which needs to be taken for further attention. Through a comprehensive analysis, this study introduces a conceptual framework for the application of e-commerce technologies in accelerating SME operation. Content analysis methodology was adopted for generating the outcome associated with the success of the technologies in SMEs.

## 1   Introduction

E-commerce technologies have proliferated over the past two decades for their growing applications. The applications are rapidly increasing in improving productivity and relevance in the online business sector [6]. E-commerce literature provides insights of various latest understanding of the effective operation of online businesses. Given that the business world is rapidly changing, it is crucial to understand the advantages matrix and different types of applications of e-commerce technologies and their appropriate meanings in terms of achieving business success, particularly for B2C businesses. Acknowledging this, in this paper we explore insights of the application of B2C e-commerce technologies in Saudi Arabian context, so that new conceptual framework of e-commerce technologies application can be outlined that would ensure the success of business.

There are numerous literature review studies in e-commerce. According to Statista [13], 90% of individuals in Saudi Arabia have access to the internet. The statistics also show that up to 80% of the population participates in online shopping. Grbovic et al. [3] examined the mandatory features advertising should have to promote e-commerce after executing information technology, considering that it is a primary tool used to boost the effectiveness of technology adoption. Hallikainen and Laukkanen [6] reviewed the influence of culture on conviction in e-commerce. Arguably, nationwide culture often unswervingly influences perceived trust. Alkhalil et al [1] Support the arguments by explaining that businesses in Saudi have identified the role of e-commerce in improving consumer trust since the technologies enhance the efficiency of the entities. Consumers in Saudi prefer to transact with entities that have effective online systems that promote their fast and effective service delivery.

## 2 Study Background

In the business world, technological advancements have impacted nearly every facet of the sector. Businesses have now adopted artificial intelligence, chatbots, and virtual assistants in the retail industry into their routine operations [9]. Although the reasons for adoption may vary from business to business, the primary motives are to help provide exceptional customer experience through instant communication, providing support without the need of data management and security, live employees, and more [15]. The primary e-commerce technologies already adopted by SME's often include electronic funds handover, record administration systems, computerized data collection schemes, internet marketing, online transaction processing, and mobile transfer [5]. For instance, in Saudi Arabia, nearly all restaurants in have adopted an electronic payment system and POS system for making orders, recording sales, and printing reports for sales, employee performance, profits, fast-moving, among others. In 2020, debit cards were the most dominant payment method for online shopping in the Kingdom, comprising 60% of all transactions [5]. Overall, most e-commerce technologies such as electronic transfer of funds, internet marketing, and digital inventory management systems have already been implemented by a significant number of SMEs in Saudi Arabia, and more growth is expected in the future [2].

The effects of e-commerce can be seen in various sectors of the Saudi economy. For instance, the revenue earned by SME's in the transport, hospitality, and food sector grew by nearly 50% in the past two years [5]. During the same duration, the government revenue attributed to these sectors grew by over 60% [5]. In a study to assess the adoption of e-commerce in the entertainment sector, Purwati [12] found that ease of payment, availability of electronic payment systems, and exceptional customer service by SMEs in Saudi's entertainment sector have facilitated the fast growth of cinemas and other entertainment joints.

**Table 1.** Critical analysis of the previous studies of e-commerce.

| Journal articles | The current literature review assessed | Articulated research gaps for the current research |
| --- | --- | --- |
| Export.gov [18] | Significant trends in e-commerce development insight into the state of it in the country and comparing it to other retailing technologies. | Basis for determining key performance indicators (KPIs) in assessing the success of e-commerce implementation. Further analysis will also provide recommendations to SMEs in Riyadh when adopting technologies. |
| Sacha Orloff [19] | Sixty-five percent of Saudi Arabia's population has access to the internet; online shoppers in the country have rapidly increased. | We noted that one of the essential factors contributing to the development of e-commerce is the ease of use. The relationship between ease of use and the success of e-commerce will be explored. |
| Grvobic et al. [3] | It examined the features of advertising needed to promote e-commerce after implementing information technology. | Analyses online advertising as one of the success factors in technology adoption. The study assesses how e-commerce enhances the efficiency of advertising on diverse online platforms. |
| Hallikainen and Laukkanen [6] | They examined how culture influences trust in e-commerce. National culture can directly influence perceived trust. | Ease of use will be used as one of the factors affecting the desire of consumers to use the service. |
| Mazzarol [10] | Examining SMEs commitment to Electronic -commerce, business, and marketing | Connection between the comfort of expenditure and the accomplishment of E-commerce will be discovered. |
| Xuhua et al. [15] | Explains effects of business-to-consumer e-commerce implementation on the economic advantage of manufacturing SMEs | Deliberate promotion as one of the success factors in technology implementation. |

## 2.1 Role of Technologies in E-commerce

While e-commerce has significantly simplified the customer buying experience, establishing an efficient e-commerce structure requires many tools and technologies combined to ensure the smooth performance of commercial transactions. Helal [7] described an investigation of the use of social media for e-commerce amongst Saudi small businesses, used a qualitative interpretive philosophical point of view. This study uses a multiple case study strategy. For Helal's research [7], four small businesses in Saudi Arabia were used and determined that traditional e-commerce currently has many obstacles that limit its distribution. Moreover, this study determined that social networks benefit small and medium enterprises in Saudi Arabia. The model she developed suggests a direct link between social capital, word of mouth, and trust in the context of small businesses in Saudi Arabia. Export.gov [18] Argues that a customer relationship management tool creates exceptional customer experience on a company's website while tracking those experiences. Adopting such a customer relationship management system and a dedicated customer support team gives a business a competitive advantage relative to other firms without customer engagement [7]. In addition, companies have started adopting methods for providing a personalized customer experience. Such marketing activities help build a loyal customer base and thereby increasing sales and profits [1]. To stay competitive in the market, businesses follow newest trends. Among the current most popular technologies used by enterprises are social media platforms such as Instagram, Facebook, Twitter, and others.

## 3  Research Methodology

The study is based on a systematic literature review that is conducted to point out the scope and prospect for diverse applications of e-commerce technologies for online businesses. We survey the literature in academic journals, books, and conference articles, having the objectives of collecting, soring, and synthesize existing studies related to applications of e-commence technologies. The surveyed sample articles focused on several applications of e-commence technologies for SMEs. The findings are leading to develop a new framework to ensure success of SME operation. We searched several databases such as IEEE, Elsevier, Springer, Association for Computing Machinery (ACM), IEEE Xplore, and other IS journals. In searching, we have used various keywords such as: "e-business" and "Saudi", "e-commence" and "Saudi" OR "developing countries", "e-commerce technologies and SME" and "e-business technologies and SMEs", "Saudi and SMEs" and "business technologies and SMEs".

Under this approach, we set two main objectives for the SLR study: gathering insights into diversified submissions of e-commerce skills and critical issues. In Saudi Arabia, the primary e-commerce technologies already adopted by SME's include electronic funds transfer, record running systems, automatic statistics collection structures, internet marketing, online transaction processing, and mobile transfer.

This study adopts content analysis methodology to generate the outcome linked to the success of technologies in SMEs. Creswell [20] Defines content analysis methodology as a technique incorporated to make valid and replicate references by coding and interpreting textual material. Overall, this study reviews relevant literature on B2C e-commerce for developing a new conceptual application framework of e-commerce technologies that facilitate the success of SMEs. Firstly, various studies addressing the meaning, implications, and benefits of e-commerce technologies on business are analyzed. This involves exploring varying options and functions of e-commerce technologies which are among the primary factors to transformational goals. Considering that a website is a primary tool in e-commerce, numerous studies highlight the features one should have. With such a website, unnecessary complexities that may hinder fast and efficient shopping.

## 4  Findings Leading to a Conceptual Framework

Notably, the business environment has widely accepted e-commerce enabling firms to promote their products and services for better results. With this growth, online commercial fields are rapidly growing, with a considerable number of traditional markets evolving to online platforms. Similarly, the growing electronic devices adoption among people has made B2C retail more pertinent. Thus, considering that e-commerce requires electronic devices to make its application more effective, and the use of such devices is on the rise, its implementation by SMEs has more significant potential.

User-generated reviews can boost its market share and sales through advertising for a company offering high-quality commodities and services. In a study to evaluate the impact of user ratings on websites, Gregory et al. [4] found that 53% of customers prefer buying from a company rated five stars and above. While e-commerce has meaningfully abridged the client's buying experience, inaugurating a well-organized e-commerce structure necessitates many tools and skills shared to ensure the smooth performance of commercial transactions.

**Table 2.** Some example studies for reviewing to develop the basis of the framework.

| Source studies | Contributions |
| --- | --- |
| Ingaldi and Ulewicz [8] | Understanding of basic e-business frameworks |
| Nguyen [11] | How various frameworks are being employed in various business model. |
| Gregory et al. [4] | Example of marketing frameworks that can be developed to facilitate e-commerce. |
| Wang and Han [17] | Overview understanding of how to implement e-commerce framework for small and medium sized enterprise |
| Yadav et al. [21] | How business website can be used to develop an e-commerce framework. |
| Tolstoy et al. [14] | Understanding the development of international e-commerce in SME's |
| Yalan and Wei [16] | Evaluating e-commerce frameworks application. |

The concept behind this project is to employ the Retail Merchant Associations (RMA). Even if the user doesn't have a standard PC or Internet connection, the RMA will assist in selling over the Internet. The RMA will establish a framework to allow them to create their e-commerce portals. Small merchants will tell the RMA whatever commodities they want to offer in his portal, and the RMA will collect the data and maintain the framework so that each merchant's goods or services can be published on his e-commerce site. Merchants will have to worry about receiving orders and fulfilling them promptly. Furthermore, RMA could benefit small businesses by providing a delivery service to their customers.

### 4.1 Framework Definition

To achieve the objectives above, we present a new e-commerce framework that allows small merchants to create their e-commerce platform with the help of the RMA. The RMA will handle all members' business. We have developed an e-commerce model geared toward merchants and backed by RMAs, in which merchants only require a cell phone and fax machine to take orders in their stores. Furthermore, the RMA would provide all of its members with the environment necessary to run a secure web server by sharing a single SSL certificate and a secure payment gateway, allowing all e-commerce portals to conduct all of their sales through the same secure payment gateway. As a result, all merchants will benefit from the RMA's partnerships with banks, while the Association's personnel will find it easier to administer the e-commerce infrastructure. For example, when a merchant has a ready order, he will send an SMS to the RMA's deliverer, instructing him to pick it up at his shop and transport it to the customer's location.

## 5  Discussion

Over the years, e-commerce adoption has widely grown in Saudi Arabia and beyond. E-commerce primarily encompasses the exchange of products and services electronically through the internet. Primary e-commerce tools include social media, websites, electronic payment systems such as online bank transfers, credit cards, debit cards, and PayPal. Based on the numerous studies reviewed, e-commerce can be differentiated from traditional businesses through its distinct features, including ease of use, increased intractability between users and business, among others. These elements form the basis of the impacts of e-commerce technologies on business. Naturally, customer satisfaction is a vital aspect that influences the success or failure of a business. With an easy-to-use interface, customers can effectively communicate with a company and shop without any challenges. This is evident in websites that are optimized so that customers can use their mobile devices to shop. Notably, the ability of e-commerce technologies to positively impact businesses to this extent is attributed to the growth and adoption of the Internet in Saudi Arabia.

The Saudi environment has a defined set of laws and policies that govern the economic operations of entities, including e-commerce. The nation enforced the e-commerce law that compels entities to respect the privacy of the personal data of their consumers. Such legal structures enhance the efficiency of transactions between business entities.

Further studies might be designed to improve the use of multiple technologies for system design that meets dynamic demands of SMEs in this digital economy. For example, how the emerging technological functions such as the Block Chain (e.g. defined in [22]) can be used for SME's data security and transparency. Smart SME application development study can be conducted through adopting the design science research [23, 24; 25]. Example applications can be aligned to Big Data Analytics solution design for enhancing data-driven decision (e.g. other industries [26, 27, 28, 29].